\documentclass[aps,pre,twocolumn,superscriptaddress,10pt,longbibliography,floatfix]{revtex4-1}
\usepackage[dvips]{graphicx}
\usepackage{amssymb,amsmath,bbm}
\usepackage[colorlinks]{hyperref}
\usepackage[colorinlistoftodos]{todonotes}

\newcommand{\mh}{\mathbf}
\newcommand{\escapetime}{\tau}

\newcommand{\position}{\boldsymbol{x}}

\newcommand{\positionP}{\position'}
\newcommand{\perturbation}{\delta}
\newcommand{\vectorperturbation}{\boldsymbol{\delta}}
\newcommand{\firsttime}{t_f}

\begin{document}

\title{Searching chaotic saddles in high dimensions}

\author{M. Sala}
\affiliation{Max Planck Institute for the Physics of Complex Systems,
N\"{o}thnizer Stra{\ss}e 38, 01187 Dresden, Germany}
\affiliation{Departamento de F\'\i sica, Universidade do Estado
  de Santa Catarina, 89219-710 Joinville, Brazil}
\author{J. C. Leit\~{a}o}
\affiliation{Max Planck Institute for the Physics of Complex Systems,
N\"{o}thnizer Stra{\ss}e 38, 01187 Dresden, Germany}
\author{E. G. Altmann}
\affiliation{Max Planck Institute for the Physics of Complex Systems,
N\"{o}thnizer Stra{\ss}e 38, 01187 Dresden, Germany}
\affiliation{School of Mathematics and Statistics, University of Sydney, NSW 2006, Australia}

\begin{abstract}
We propose new methods to numerically approximate non-attracting sets governing transiently-chaotic systems. Trajectories starting in a
vicinity $\Omega$ of these sets escape $\Omega$
in a finite time $\tau$ and the problem is to find initial conditions $\position \in \Omega$  with increasingly large $\tau= \tau(\position)$. We
search points $\position'$ with $\tau(\positionP)>\tau(\position)$ in a {\it search domain} in $\Omega$. Our first method considers a search
domain with size that decreases exponentially in $\tau$, with an exponent proportional to
the largest Lyapunov exponent $\lambda_1$. Our second method considers anisotropic
search domains in the {\it tangent} unstable manifold, where each direction scale as the inverse of the corresponding {\it expanding} singular value of the Jacobian matrix of the iterated map. We show that both methods outperform the state-of-the-art {\it Stagger-and-Step} method (Sweet, Nusse, and York, Phys.
Rev. Lett. {\bf 86}, 2261, 2001) but that only the anisotropic method achieves an efficiency independent of $\tau$ for the case of
high-dimensional systems with multiple positive Lyapunov exponents.  
We perform simulations in a chain of coupled H\'enon maps in up to 24 dimensions ($12$ positive Lyapunov exponents). This suggests the possibility of characterizing also non-attracting sets in spatio-temporal systems.
\end{abstract}

\maketitle

{\bf
\noindent
Maximizing functions is a traditional computational problem with numerous applications in Physics.
One of the main difficulties in this problem arises when the function has
multiple local maxima. An extreme example of this problem is found in transiently chaotic
systems because there are infinitely many points $\position$ for which the escape time $\tau(\position) \rightarrow \infty$ (stable
manifold of a saddle), but these points are distributed in phase space over a fractal set with zero volume. The difficulty is enhanced when multiple positive-Lyapunov exponents exist (hyperchaos). 
In this paper we show how we can take advantage of the regularity of chaotic systems to design an algorithm that is able to efficiently approximate the invariant manifolds of high-dimensional transiently-chaotic systems.
}

\section{\label{sec:int}Introduction}

We consider chaotic dynamical systems $F: \position_n \mapsto \position_{n+1}$ for which almost all trajectories $\position$ starting in a region of interest
$\Gamma \in R^d$ (restraining region) leave it after a finite number of iterations
$n=\escapetime$ (escape time). The zero-measure non-attracting set of trajectories that
never escape ($\tau \rightarrow \infty$) build a fractal set, the stable manifold of the
chaotic saddle that governs the asymptotic dynamics of the system. This is the familiar
picture of transient chaos, which describes a variety of physical applications~\cite{LaiTamasBook,Tel2008,Rempel2007,Taylor2007}.

An important problem in numerical investigations is to approximate the invariant sets, which amounts to maximize the escape time
$\escapetime(\position)$  (a function $\mathbbm{R}^d \mapsto \mathbbm{R}$)\footnote{The higher the escape time, the nearer the point is to the stable manifold of the invariant set, the nearer $F^{\escapetime/2}(\position)$ is to the chaotic saddle and nearer $F^{\escapetime}(\position)$ is to the unstable manifold.~\cite{LaiTamasBook}}.
Two difficulties appear in (hyperbolic) chaotic systems:
(i) the phase-space volume of trajectories with $\escapetime$ decays as $P(\escapetime) \sim e^{-\kappa \escapetime}$, where $\kappa$ is the escape rate of the system;
(ii) the set of points for which $\escapetime\rightarrow \infty$ is a complicated fractal set with $D<d$ (the stable manifold of the chaotic saddle).
Different numerical methods have been designed to approximate the invariant sets~\cite{Sweet2001, Bollt2005} or to compute specific properties such as the
fractal dimension~\cite{DeMoura2001}. While the success of these methods for systems with a single positive Lyapunov exponent ($\lambda_1 >0,
\lambda_{2,\ldots, d} <0)$ is well established, for
the challenging case of hyperchaotic systems the most popular~\cite{LaiTamasBook,Taylor2007} and
efficient solution is the
Stagger-and-Step method proposed by Sweet and Yorke in 2001~\cite{Sweet2001}. Here we revisit the core problem of all these methods, which
can be formulated as follows: starting from a point $\position$ with escape-time $\escapetime$, the task is to search a new point $\positionP$ with higher
escape-time $\escapetime(\positionP)>\escapetime(\position)$. 
A simple search strategy is to consider the pre-image of $\position$, $\positionP = F^{-1}(\position)$, which by construction has $\escapetime(\positionP)=\escapetime(\position)+1$~\cite{Dellago2002,Leitao2014}.
This strategy is, however, of limited use because repeating it quickly leads $\positionP$ to fall outside $\Gamma$ (the search domain).
The other generic search strategy, which we focus in this paper, is to search in a vicinity $\vectorperturbation(\position)$ of the original point $\positionP=\position+\vectorperturbation(\position)$.

In this paper we investigate the efficiency of different search strategies (choices of $\vectorperturbation(\position)$) and compare their efficiency as the number of searches needed in order to find a trajectory with a pre-assigned escape time $\escapetime=\escapetime_{max}$.
%
We start introducing the system in which we test our algorithms (in Sec.~\ref{sec.Henon}).
The first method we test is the stagger and step (in Sec.~\ref{sec.stagger}).
We conclude that its efficiency grows faster than linearly with $\escapetime_{max}$ due to an inefficient search strategy $\vectorperturbation(\position)$.
To address this problem, we propose (in Sec.~\ref{sec.adaptive}) an adaptive method which considers a $\vectorperturbation(\position)$ whose size decays with $\escapetime(\position)$.
A comparison of the efficiency of the two methods (in Sec.~\ref{sec.efficiency}) confirms the benefit of our adaptive proposal but shows that the existence of more than one unstable direction in the phase space poses a major limitation to both methods.
We show (in Sec.~\ref{sec.anisotropic}) that this limitation can be overcome using a specific anisotropic distribution of $\vectorperturbation(\position)$.
Finally, we discuss the implications of our results (in Sec.~\ref{sec.conclusions}).

\section{Coupled H\'enon maps}\label{sec.Henon}

We test different methods through simulations in a family of dissipative systems with
variable dimension that shows transient chaos.
We consider a $d=2N$ dimensional map $F: (x_i,y_i)_n \mapsto (x_i, y_i)_{n+1}$
built coupling $i=1, \ldots, N$ two-dimensional maps 
\begin{equation}
\left(\begin{array}{c}
x_{i}\\
y_{i}
\end{array}\right)_{n+1} = 
\left(\begin{array}{c}
A_i - x_i^2 + B y_i + k(x _i - x_{i+1})\\
x_i
\end{array}\right)_n,
\label{eq.henon}
\end{equation}
with $i=1, \ldots, N$, periodic boundary conditions $(N+1 \equiv 1)$, $k=0.4 ,B=0.3,
A_1=3$ (if $N>1$), $A_N=5,$ and $A_i=A_1+(A_N-A_1)(i-1)/(N-1)$.  Initial conditions are
chosen on a $2N$ hypercube $\Gamma=[-4,4]^{2N}$ (restraining region), which according to our exploratory
numerical simulations contains the invariant
set of the system for all $N$. This system was defined
in Ref.~\cite{Leitao2013}, recovers for $N=1$ a well-studied chaotic H\'enon map, and recovers for
$N=2$ the  system used in the original paper of the Stagger-and-Step method~\cite{Sweet2001}. For
increasing dimensionality~$N$, our numerical simulations systematically show exponential
decay in time of the number of surviving trajectories,  $N$ positive Lyapunov exponents,
and fractal invariant sets.

\section{Stagger method\label{sec.stagger}}

The state-of-the-art for computing approximations of the chaotic invariant sets in systems
with more than one $\lambda_i >0$  is the Stagger-and-Step method\cite{Sweet2001}.
It combines two procedures: a random search for points with high escape-time (the stagger part), followed by the construction of pseudo-orbits starting from such points (the step part).
Here we are interested in testing and improving the stagger part (the step part can be incorporated in the methods discussed below).
The stagger search method considers $\vectorperturbation=\perturbation\hat{\mh{u}}$, where $\hat{\mh{u}}$ is a unit vector chosen with uniform probability in
the unit sphere (isotropic) and $\perturbation$ is chosen randomly from a truncated scale-free distribution given by
\begin{equation}
\label{eq.stagger}
P(\perturbation)= \frac{Z}{\perturbation} \text{ for } \perturbation_{min} < \perturbation < \perturbation_{max},
\end{equation}
where $Z$ is the normalization and where we use $\perturbation_{max}$ to be of the system size and $\perturbation_{min}$ to be equal to the (potentially variable) machine precision
\footnote{In Ref.~\cite{Sweet2001} the distribution is constructed by considering $\perturbation=10^{-s}$ with $s$ chosen randomly from a uniform distribution in $[\log_{10}(\perturbation_{min}), \log_{10}(\perturbation_{max})]$}.
Starting from a point $\position$ with $\tau(\position)$, the method consists in:
\begin{itemize}
\item[1.] calculate a proposal position $\positionP=\position+\vectorperturbation$ and respective escape time
  $\escapetime(\positionP)$, where $\vectorperturbation = \delta \hat{\mh{u}}$ and $\delta$ is drawn randomly from Eq.~(\ref{eq.stagger});
\item[2.] accept the proposal if $\escapetime(\positionP) > \escapetime(\position)$ (a
  success) or stay in the same position $\position$ if not;
\item[3.] go to 1 until $\escapetime(\position)$ reaches a maximum escape time
  $\escapetime_{\max}$.
\end{itemize}

\begin{figure}[bt!]
\centering
\includegraphics[width=\columnwidth]{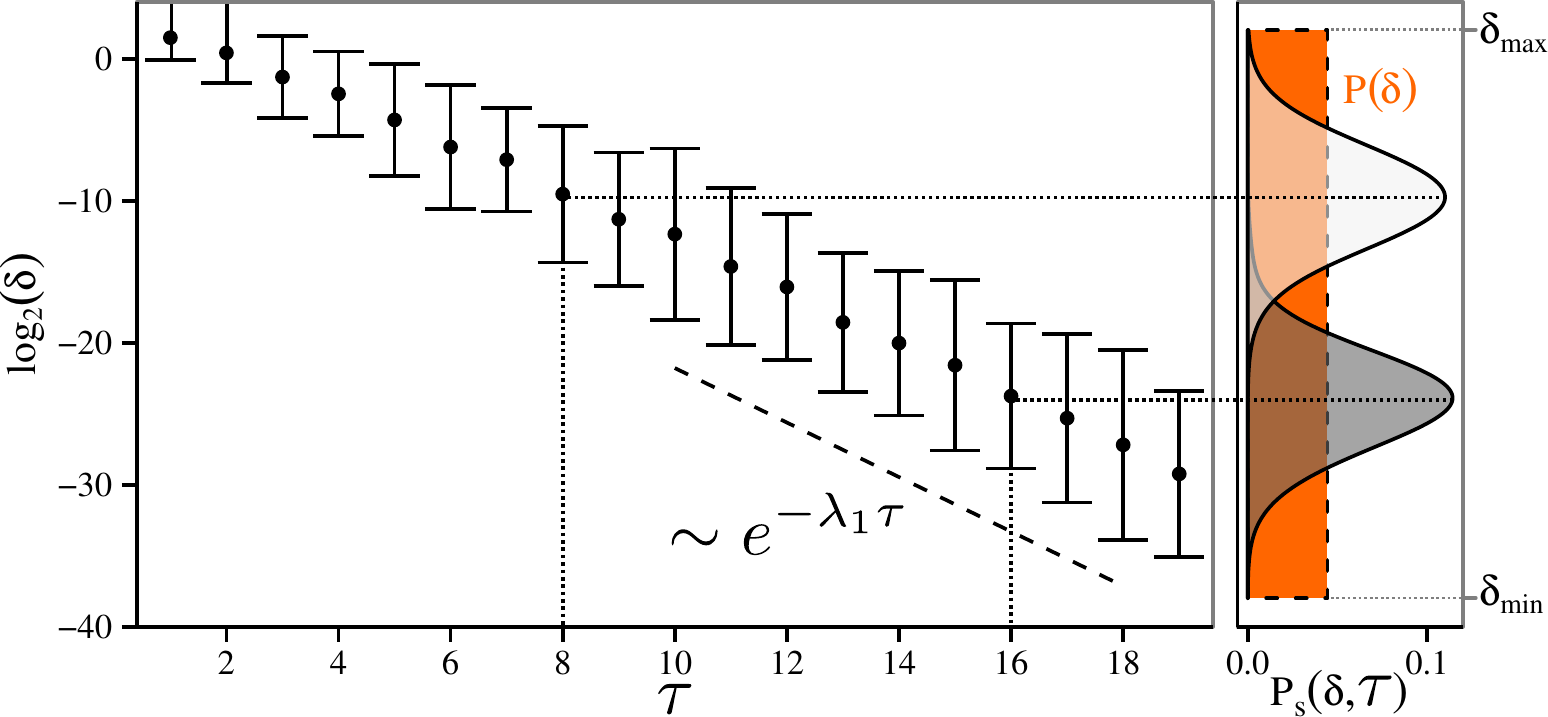}
\caption{ The successful proposals $\positionP=\position+\vectorperturbation$ of the Stagger method depend on the escape time $\escapetime=\escapetime(x)$.
Left: average and standard deviation of the length of successful proposed values
$\perturbation=|\vectorperturbation|$ for a given $\escapetime(\position)$ (i.e., for which
$\positionP=\position+\vectorperturbation$
leads to $\escapetime(\positionP)>\escapetime(\position)$).
The dashed line corresponds to an exponential decay with an exponent $\lambda_1=1.33$, 
the largest Lyapunov exponent of this system as reported in Ref.~\cite{Sweet2001}.
Right: the distributions $P_S(\delta)$ of successful $\delta$ for two different
$\escapetime$'s (8 in white and 16 in gray) along with the Stagger distribution $P(\delta)$ defined in Eq.~(\ref{eq.stagger}).
stagger.
Results are obtained from independent simulations in system~(\ref{eq.henon}) with $N=2$, $\tau_{\max}=19, \delta_{min}=2^{-39}$, and $\delta_{\max}=1$.
}
\label{fig.I}
\end{figure}

In Fig.~\ref{fig.I} we test the Stagger proposal~(\ref{eq.stagger}) by investigating how the successful proposals depend on $\escapetime$. 
We find that, on average, the magnitude of the successful proposal $\perturbation^*$ decays with $\escapetime$, consistent with
\begin{equation}\label{eq.rlambda}
\perturbation^* \sim e^{-\lambda_1 \escapetime},
\end{equation}
where $\lambda_1$ is the largest Lyapunov exponent of the system. This is justified by the notion that the points that escape at time $\escapetime$ are the $\escapetime$-th pre-images of a suitably defined escape region, and thus the size of its pre-images decrease exponentially with the maximum Lyapunov exponent~\cite{Grunwald2008,Leitao2013}.
The stagger proposal~(\ref{eq.stagger}) is independent of $\escapetime$ and does not
explore this regularity (right panel of Fig.~\ref{fig.I}): the overlap is small between
the perturbation range $[\perturbation_{min}, \perturbation_{max}]$ and the range of
successful perturbations whose average scales with Eq.~(\ref{eq.rlambda}).
As $\escapetime_{max}$ increases, in order to find successful trajectories one has to
reduce $\perturbation_{\min}$ (e.g., by increasing the machine precision).
This increases the search range $[\log \delta_{\min}, \log \delta_{\max}]$, which further
decreases the overlap with the range of successful proposals. As a consequence, the efficiency of the stagger search decreases with $\tau_{\max}$.
The natural question we explore in the next section is how to use the regularity in the
$\delta$ vs. $\tau$ relation -- Eq.~(\ref{eq.rlambda}) -- to design a more efficient search
algorithm (even if $\lambda_1$ is unknown).

\section{Adaptive method\label{sec.adaptive}}

Motivated by the idea that there is a typical scale $\delta$ for a successful search at each value of $\delta$, we
propose a search algorithm using the same procedure as the Stagger, but with a
distribution $P(\perturbation)$ different from Eq.~(\ref{eq.stagger}). Specifically, we
consider a normal distribution with average 0 and standard deviation
$\sigma(\position)$. We choose this distribution because the distance $|\position -
\positionP|$ is a half-normal distribution which has expected value proportional to
$\sigma$ (the characteristic length of proposals). Ideally, we would choose $\sigma\sim
\delta$ to scale as $\sigma(\escapetime) \sim e^{-\lambda_1 \escapetime}$. However, since
$\lambda_1$ is typically unknown, we use the adaptive procedure proposed in Ref.~\cite{Leitao2013}, where $\sigma$ is adapted at each step of the simulation by
\begin{equation}\label{eq.sigma}
\sigma_{t+1}(\escapetime) = \left\{\begin{array}{c}
\sigma_t(\escapetime)/f   \text{ for } \escapetime(\positionP) < \escapetime(\position) \\
\sigma_t(\escapetime) f  \text{ for } \escapetime(\positionP) = \escapetime(\position),
\end{array}\right.
\end{equation}
where $f\gtrapprox 1$ is a free parameter (we use $f=1.1$).
The idea behind this procedure is that a too large $\sigma$ will propose points $\positionP$ that are mostly drawn from the escape time distribution, which will have $\escapetime(\positionP) \approx 1$ and thus are mostly unsuccessful.
Thus, we decrease $\sigma$ when a low escape time is proposed.
Likewise, when $\sigma$ is too small the points $\positionP$ will be indistinguishable from $\position$ after $\escapetime$ iterations and thus $\escapetime(\positionP)=\escapetime(\position)$.
In this case we increase $\sigma$.
We expect this iterative procedure to converge to a value of $\sigma$ for which the
desired $\escapetime(\positionP)$ is achieved. In order to increase the mixing in
$\position$,  we accept proposals
$\positionP$ for which $\tau(\positionP)=\tau(\position)$ (differently from the Stagger method summarized above). 

\begin{figure}[bt!]
\centering
\includegraphics[width=\columnwidth]{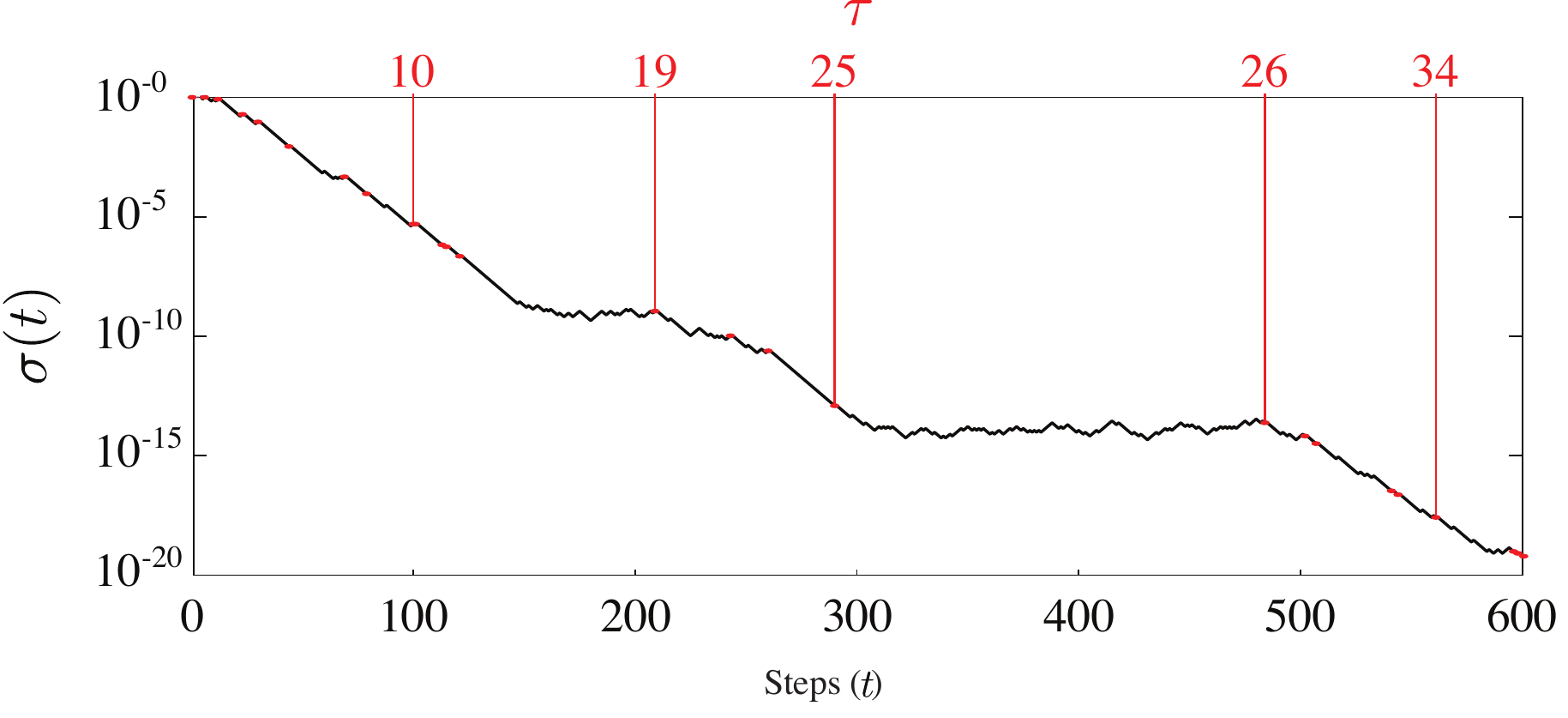}
\includegraphics[width=\columnwidth]{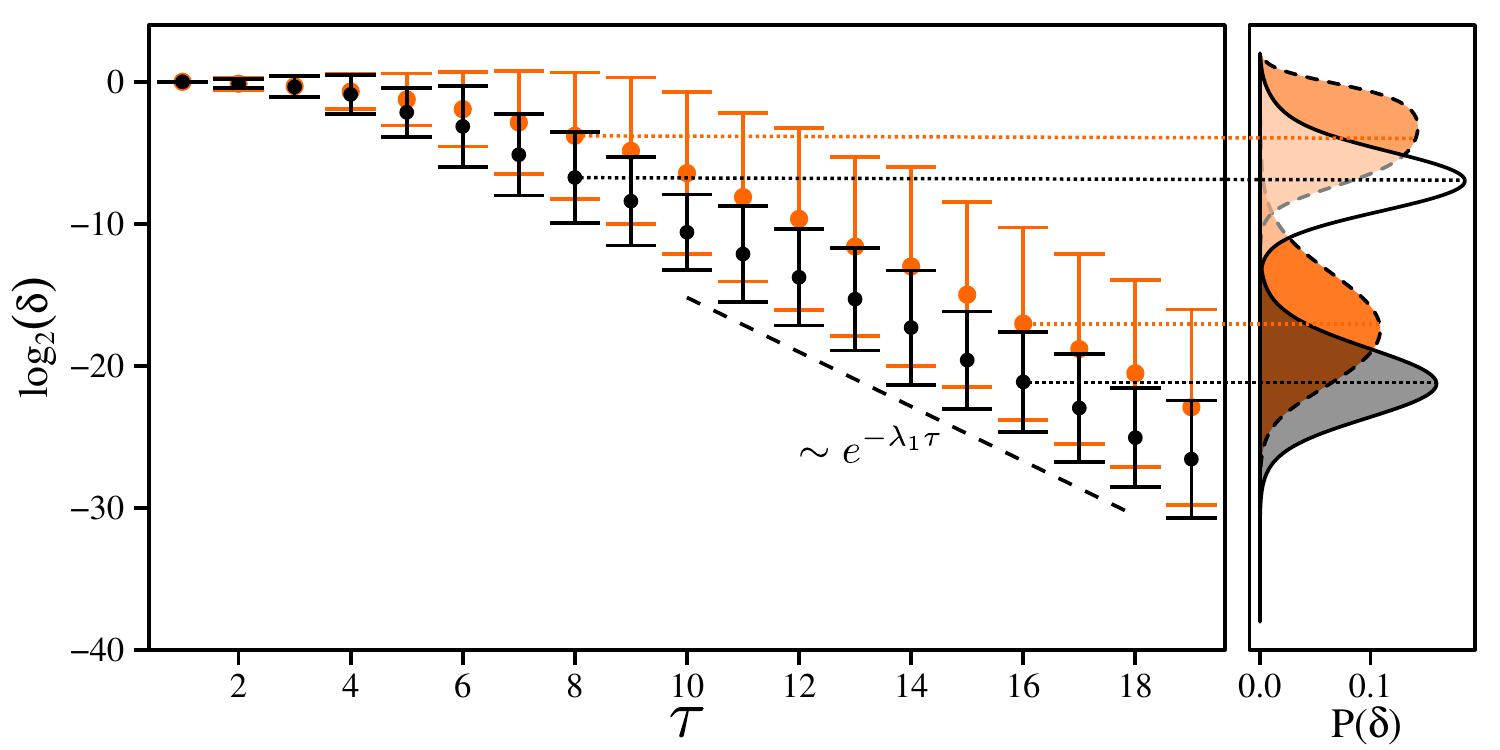}
\caption{The adaptive method of proposal.
Upper panel: evolution of a single realization of the adaptive method as a function of the number
of searches $t$ (Monte Carlo time). The line shows the evolution of the scale $\sigma$ of the proposal
according to Eq.~(\ref{eq.sigma}). Each red dot represents a step on which the escape
time $\tau$ increased (a success). 
Lower panel: average and standard deviation of the proposals (blue) and successful
proposals (black) computed over $100$ independent simulations (compare to Fig.~\ref{fig.I}). Left: in blue is the average and standard deviation
displacement proposed for each $\escapetime$, averaged over 100 independent
simulations. Right: full distributions for $\escapetime=8$ and $\escapetime=16$. White
($\tau=8$) and gray $(\tau=16)$ represent the success, light blue 
($\escapetime=8$) and blue ($\escapetime=16$) represent the proposals. 
Simulations were performed in system~(\ref{eq.henon}) with $N=2$ and using $f=1.1$.
}
\label{fig.II}
\end{figure}

Our numerical simulations are summarized in Fig.~\ref{fig.II} and indicate that the
adaptive method with Eq.~(\ref{eq.sigma}) converges to proposals with the appropriate
$\sigma$. In particular, both the distributions of the proposed and successful
distances $\delta$ scale with $\tau$ as $ \sim e^{-\lambda_1 \escapetime}$,
in agreement with Eq.~(\ref{eq.rlambda}).
The advantage of this procedure in comparison to the Stagger is that the proposal at a
given $\tau$ does not depend on $\tau_{\max}$ and therefore the efficiency does not decay
with $\tau_{\max}$.
In the next section we compare the efficiency of the adaptive and Stagger method as a function of $\tau$ for a fixed $\tau_{\max}$.

\section{Efficiency \label{sec.efficiency}}

We quantify the efficiency of a search method by the rate of success in finding
trajectories.
We define the $\tau$-dependent {\it success rate} $C(\escapetime)$ as an average $\langle \ldots \rangle$ over the different realizations $r$ of the algorithm as
\begin{equation}\label{eq.c}
C(\escapetime) \equiv \langle C_r(\escapetime) \rangle = \left\langle \frac{\firsttime(\escapetime)}{\escapetime}\right\rangle ,
\end{equation}
where $\firsttime(\escapetime)$ is the total number of proposals needed for the realization $r$ to reach $\tau$ (e.g., in the upper panel of Fig.~\ref{fig.II}, $C_r(\escapetime=26)=490/26$ because the algorithm needed $490$ search steps to reach $\tau=26$)
\footnote{To compute $\escapetime$, the map has to be iterated $\escapetime$ times and thus the number of map iterations needed to reach $\tau_{\max}$ is  of the order of $\sum_{\tau=1}^{\tau_{\max}} C(\escapetime)\escapetime$.}.
For $\escapetime \gg 1$, the escape time function is self similar with $\escapetime$ and thus we aim for the algorithm to be independent of $\escapetime$.
This optimal situation corresponds to have $C(\escapetime\rightarrow \infty) = C_r(\escapetime\rightarrow \infty)\equiv C$ where $C$ is independent of $\tau$ 
that thus can be used to characterize the efficiency of the method in the particular system.

\begin{figure*}
\includegraphics[width=1.4\columnwidth]{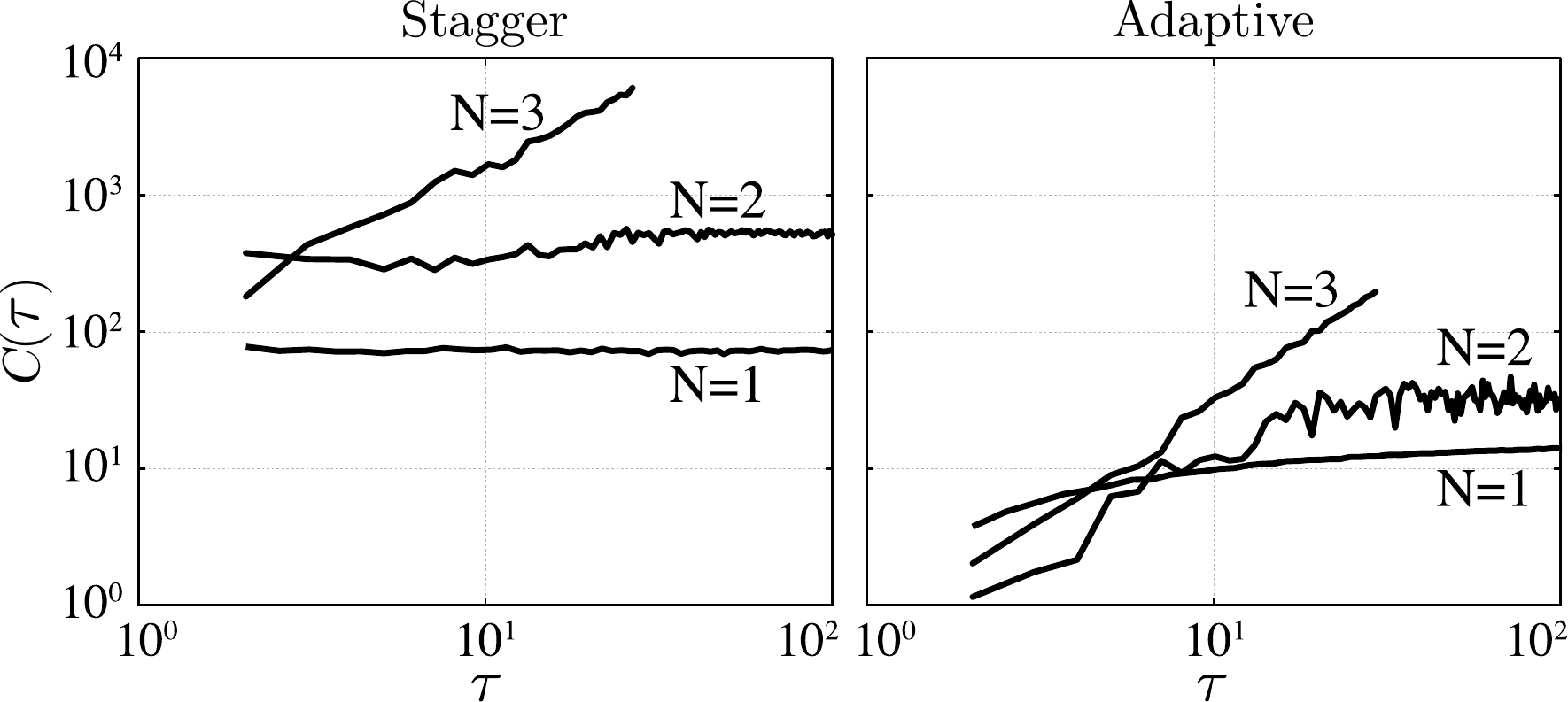}
\caption{Efficiency of the Stagger and Adaptive methods depend on the dimension $N$.
Success rate $C(\escapetime)$ as a function of $\escapetime$ for the Stagger (left) and adaptive (right) methods, for $N=1,...,4$.
Each curve corresponds to an average over $100$ independent runs starting at $\escapetime_{\min}=1$ and ending at $\escapetime_{\max}=32$ or $\escapetime_{\max}=64$.
}
\label{fig.III}
\end{figure*}

In Fig.~\ref{fig.III} we present the numerical results for
$C(\escapetime)$ for the Stagger and Adaptive methods.
For the (easy) case of low dimensions ($N=1$ and $N=2$), we see a convergence in $\tau$ to the well-defined success rate $C$.
The value of $C$ for the adaptive method is lower than for the Stagger method,
indicating that it is more efficient. The $C$ value of the Stagger could be reduced by
increasing $\delta_{\min}$, at the expense of reducing $\tau_{\max}$ (the maximum obtainable $\tau$).
More surprising, the (difficult) case of high dimensions $N$ shows that neither the
Stagger nor the Adaptive method have a constant cost $C(\escapetime)$.
This suggests that, contrary to previous claims~\cite{Sweet2001, LaiTamasBook}, the Stagger method becomes inefficient in higher dimensions (it shows a qualitatively different scaling).
We verified that if the proposals are drawn from a Gaussian with width scaling as in Eq.~(\ref{eq.rlambda}), $C(\tau)$ also grows with $\tau$ for $N=3$.
Overall, our results hint that the Stagger method may not be as robust to increasing dimensions as suggested in Refs.~\cite{Sweet2001, LaiTamasBook}.

\section{Anisotropic method\label{sec.anisotropic}}

The results above indicate that a choice of the right scale of the proposal $P(\delta)$ is not sufficient to efficiently
search in high-dimensions. The fact that the problems arise (become more severe) when more
than one Lyapunov exponent is positive suggests that it is related to the existence of different stretching rates along different directions in the phase space. Based on this insight,
we consider in this section $\vectorperturbation$ drawn from a non-isotropic distribution
(i.e., $\vectorperturbation \neq \delta \hat{\mh{u}}$). 
Ref.~\cite{Bollt2005} already employed an anisotropic search in high dimensions, considering a deterministic (gradient) search in a single direction.
Our approach here is to generalize our ideas to an anisotropic search.

Consider a point $\position$ with an escape time $\escapetime(\position) = \escapetime$ and its $\escapetime$-th evolution in time, $\position_\escapetime = \boldsymbol{F}^\escapetime(\position)$.
An initial infinitesimal displacement $d\position$ close to $\position$ will, after a time $\escapetime$, be transformed into $d\position_\escapetime$ according to the dynamics in the tangent space
\begin{equation}
d\position_\escapetime =J(\position) d\position,
\label{eq:tangent}
\end{equation}
where $J(\position)\equiv d\boldsymbol{F}^\escapetime(\position)/d\position$ is the Jacobian matrix of the map iterated $\escapetime$ times.
In an isotropic proposal, $\vectorperturbation$ is proposed isotropically around $\position$ (the previous sections showed that the optimal scale for $|\vectorperturbation|$ scales with the largest Lyapunov exponent $\lambda_1$).
The problem with isotropic proposals is that when the distribution of $d\position$ is isotropic, the distribution of $d\position_\escapetime$ is not: it contracts and stretches along the singular/covariant basis according, respectively, to the singular/Lyapunov spectrum of matrix ${\bf J}$. In particular, for large $\escapetime$, $d\position_\escapetime$ is aligned with the most unstable direction (associated to $\lambda_1$).
This poses a major problem because this direction (which is a one dimensional object) may not intersect any region of the phase-space with higher escape time for a given point $\position$, and this situation makes the algorithm stuck on that point.
We expect this to become more dramatic with increasing dimension as the one dimensional search volume becomes increasingly small in comparison to the total phase-space dimension D.

\begin{figure}[bt!]
\centering
\includegraphics[width=\columnwidth]{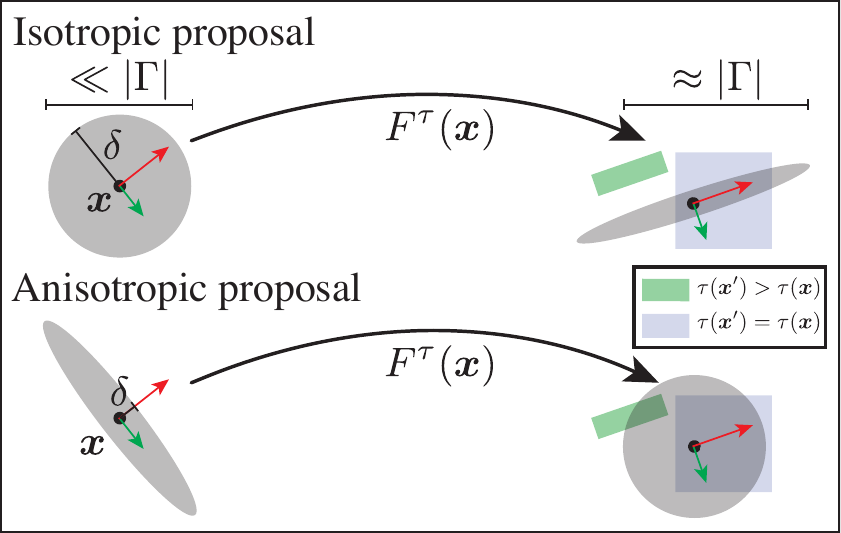}
\caption{
Anisotropic proposals in $\position$ lead to isotropic points in $\position_\tau=F^\tau(\position)$.
The figure represents two stretching directions of a phase space (red arrow associated to $\lambda_1$, green arrow, associated to $\lambda_2$). Proposals (gray regions on the left) and its corresponding region after $\escapetime = \escapetime(\position)$ iterations are shown. 
After $\escapetime$ iterations, $\position_\tau$ ($\bullet$) is in the exit set (blue region). The isotropic proposal fails when the target region with $\tau' > \tau$  (green rectangle) does not intersect the one-dimensional most-unstable direction.
Our anisotropic proposal in $\position$ is constructed so that it is isotropic in $\position_\tau\equiv F^\tau(\position)$. This maximizes the search dimension thus improving the probability to find higher escape times.
}
\label{fig.explanation}
\end{figure}

\begin{figure*}
\centering
\includegraphics[width=\textwidth]{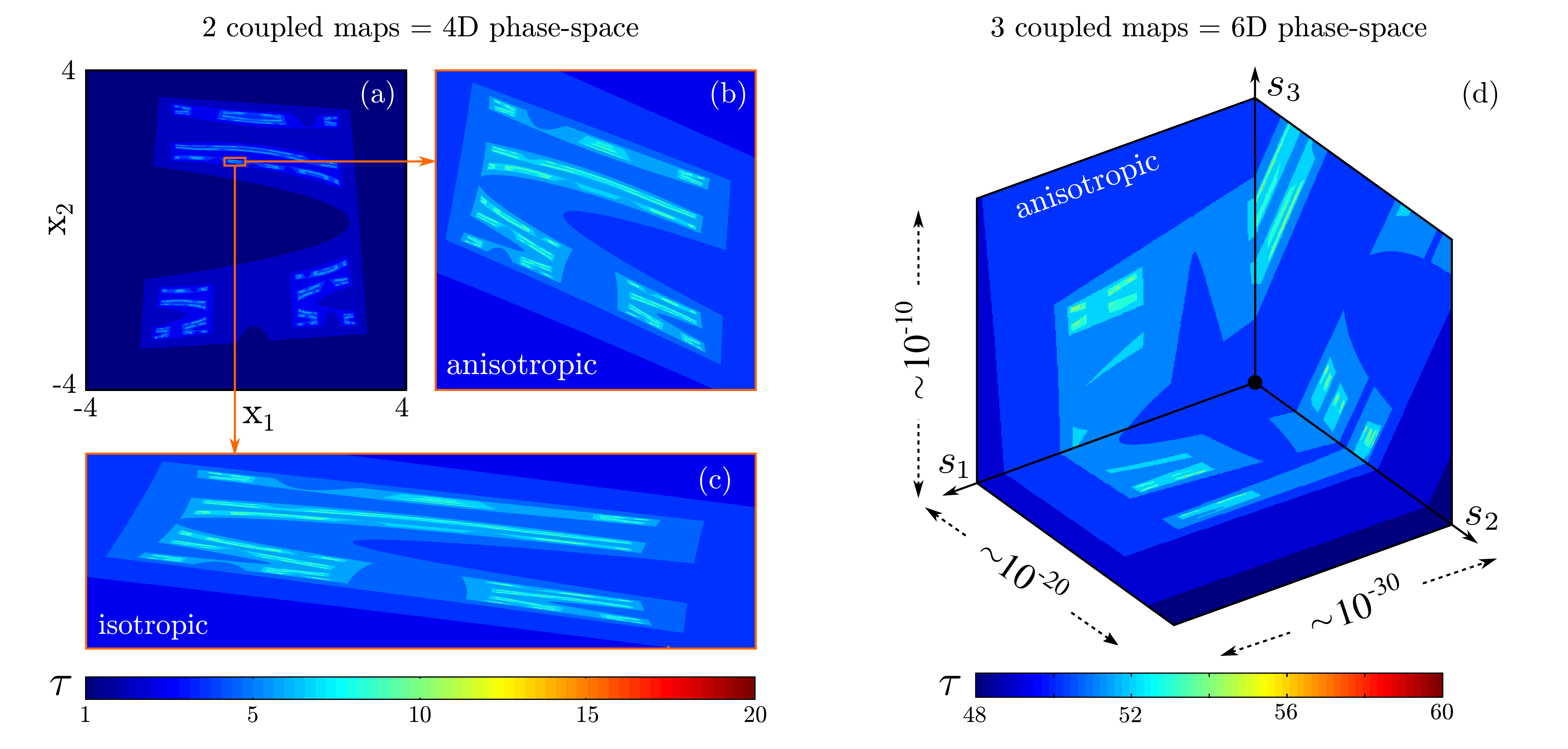}
\caption{A self-similar distribution of escape times~$\tau$ is observed through anistropic magnifications of the phase space of system~(\ref{eq.henon}). Starting from the $x_1,x_2$ cut of the phase space for $N=2$ -- in panel (a) --, we obtain a similar (rotated) distribution after performing an anisotropic magnification -- in panel (b) -- while the isotropic magnification leads to a distorted picture -- in panel (c). A similar picture is seen also at a tiny portion of the phase space for $N=3$ after an anisotropic magnification, as shown in panel (d).}
\label{fig.5}
\end{figure*}

Our main idea is to use an anisotropic perturbation $\vectorperturbation$ such that, by construction, its evolution after $\escapetime$ iterations, $\perturbation \position_\escapetime$, is isotropic on the unstable directions, see Fig.~\ref{fig.explanation}.
We can use Eq.~(\ref{eq:tangent}) to construct such perturbation because, for a small perturbation $\perturbation \position$, we have
\begin{equation}
\perturbation \position = J^{-1} \perturbation \position_\escapetime.
\label{eq:anisotropic_bad}
\end{equation}
However, we should not use ${\bf J}^{-1}$ directly because it contains a mix of unstable directions and stable directions.
Specifically, on its single value decomposition, ${\bf J}={\bf U\Sigma V}^T$ where ${\bf U}$ and ${\bf V}$ are unitary matrices and ${\bf \Sigma}$ is a diagonal matrix, ${\bf \Sigma}$ contains entries higher than 1 (corresponding to unstable directions), and smaller than 1 (corresponding to stable directions).
Because ${\bf J}^{-1} = {\bf V\Sigma}^{-1} {\bf U}^T$, Eq.~(\ref{eq:anisotropic_bad}) would generate a proposal $\perturbation \position$ with extremely large values due to the diagonal values higher than 1 of ${\bf \Sigma}^{-1}$ (corresponding to the stable directions), which would be a problem because $\perturbation \position$ would violate the approximation that $\perturbation \position$ is small.
In order to obtain an isotropic proposal along all unstable directions of the $\escapetime$-th iteration of the map, we use a transformed matrix ${\bf J}^{-1}$ that considers only the unstable directions. This is achieved as
\begin{equation}
\perturbation \position = {\bf V}({\bf \Sigma}^+)^{-1} {\bf U}^T \perturbation \position_\escapetime,
\label{eq:anisotropic}
\end{equation}
where the diagonal matrix ${\bf \Sigma}^+$ is constructed by setting to zero the entries smaller than 1 in ${\bf \Sigma}$. 
The diagonal elements of ${\bf \Sigma}^+$ quantify the strength of the divergence along the directions defined by the columns of $U$ on forward iterations.
However, they are not related to the (finite-time) Lyapunov exponents via $-\log(({\bf \Sigma}^+)_{ii})/\escapetime$.
A good approximation of the (finite-time $\tau_{cs}$) Lyapunov exponents and covariant Lyapunov vectors is obtained storing the Jacobian matrix of a piece of the trajectory in the interval $[\escapetime/2 - \tau_{cs}, \escapetime/2 + \tau_{cs}]$~\cite{LaiTamasBook,Note1}. By finding long escape-times $\tau$, our method allow us to select large $\tau_{cs}$ (i.e., segments of orbit that
stay inside the system {\it long enough}) to allow for the computation of the spectrum of Lyapunov
exponents and their associated covariant vectors~\cite{Kuptsov2012}. 

\begin{figure}
\centering
\includegraphics[width=\columnwidth]{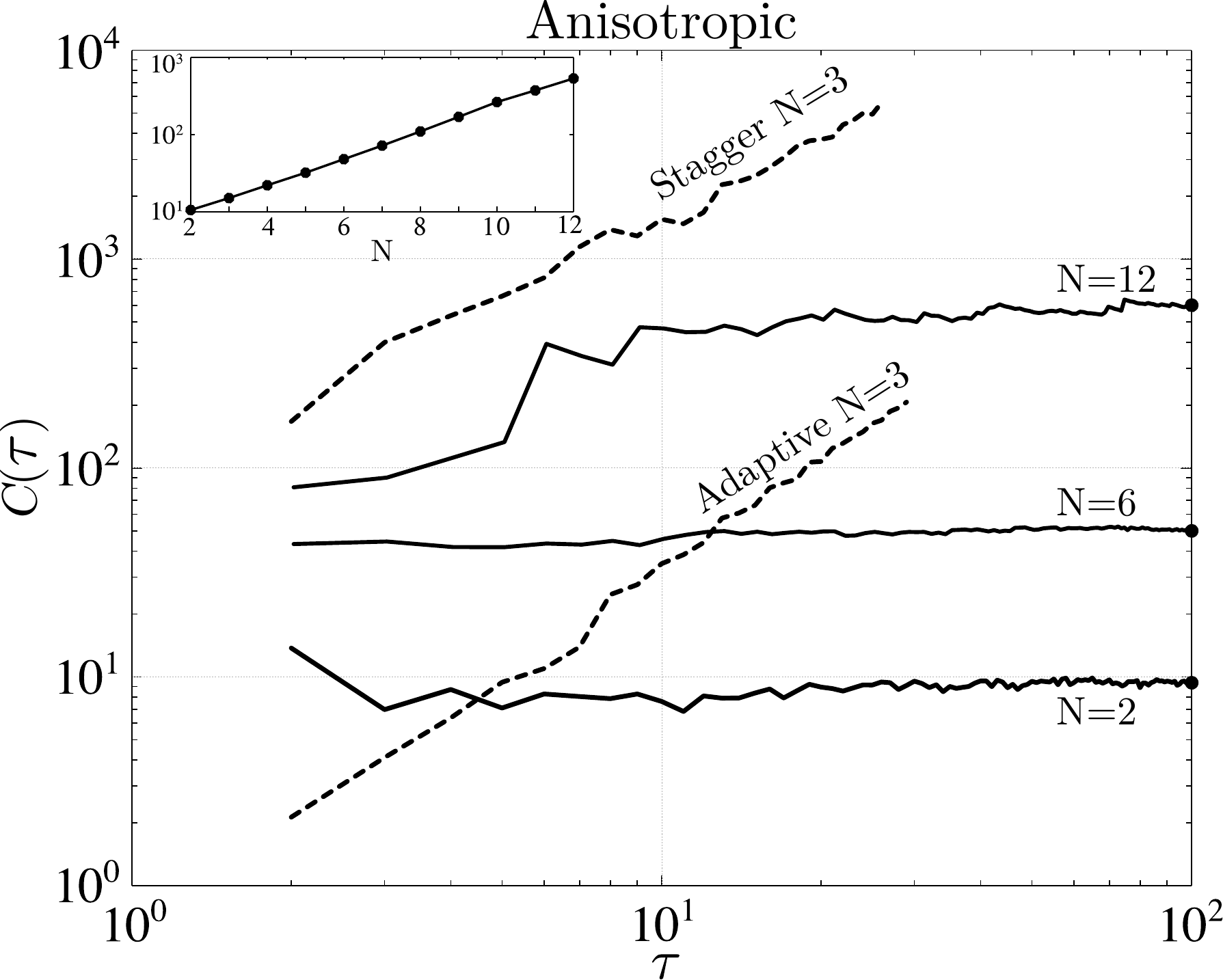}
\caption{The cost of the anisotropic method does not increase with $\tau$.
(a) The average cost, computed with the same parameters as in Fig. 3, for the anisotropic method (blue) for 3 different phase-space dimensions, stagger and adaptive for 6D.
Inset shows how the method scales with increasing phase-space dimension.}
\label{fig.IV}
\end{figure}

A comparison between isotropic and anisotropic magnification of the phase space of system~(\ref{eq.henon}) appears in Fig.~\ref{fig.5}. The efficiency of our numerical simulations using the anisotropic method is summarized in Fig.~\ref{fig.IV}.
They confirm that the cost $C(\escapetime)$ is independent of $\escapetime$ for up to $N=12$, and that it is more efficient than the stagger and the isotropic proposal.
It also indicates that the cost increases exponentially with increasing dimension, a consequence of the increasing search space.
This method also requires computing the product of Jacobian matrices and a single-value decomposition, both of which have a cost that increases with $d$.
Nevertheless, our method allows to find trajectories with very high escape times using lower computational resources than the previous approaches.

\section{Summary and  Conclusions\label{sec.conclusions}}

Chaotic saddles are a generic feature of open non-linear dynamical systems that are expected to appear in arbitrary large dimensions. For instance, they are known to appear in high-dimensional systems such as delay differential equations~\cite{Taylor2007} and spatially-extended systems~\cite{Rempel2007,Tel2008}.
Finding long-living trajectories that approximate such chaotic saddles is a critical requirement to numerically study them.
The computational difficulty in this task is that the escape-time function $\tau(x)$ is extremely rough and that the phase-space volume with escape time $\tau'>\tau$ decays exponentially with $\tau$.

The state-of-the-art method is the stagger method, that searches trajectories using an isotropic and power-law distributed proposal, intended to cover all the potential scales of the escape-time function.
For low-dimensional systems, we showed that the stagger method can be improved by using a proposal distribution  with a $\tau$-dependent length scale (governed by the largest Lyapunov exponent, Eq.~(\ref{eq.rlambda})). We showed also that an adaptive procedure can be used when the Lyapunov exponent is unknown (Eq.~(\ref{eq.sigma})).
For high-dimensional systems, we found that both the stagger and the adaptive methods become increasingly inefficient for increasing $\tau$.
We presented an explanation of why this is so, based on general arguments of dimensionality and dynamics of the tangent space. This motivated the main result of our paper: the introduction of an anisotropic proposal distribution that considerably improves the efficiency of searches of high-dimensional chaotic saddles.

%
We expect our results to bring new insights in the study of chaotic saddles in high-dimensional (spatially-extended) systems and also to be applicable in cases beyond the ones discussed above.  
For instance, proposals increasing the escape time are required also in the more general problem of sampling the phase space of open chaotic systems~\cite{Leitao2013}. 
Our results and methods remain valid also in continuous-time systems: while the application of the adaptive method is straightforward, for the anisotropic method our singular-value-based filter (used to construct Eq.~(\ref{eq:anisotropic})) has to be implemented on the Jacobian matrix of the set of ordinary differential equations (see, e.g., Ref.~\cite{Kuptsov2012}).

\section*{Acknowledgements}
MS was supported by CAPES (Brazil). JCL was supported by FCT (Portugal), grant no. SFRH/BD/90050/2012. EGA thanks T. T\'el for helpful discussions.

\bibliographystyle{apsrev4-1}

\begin{thebibliography}{15}%
\makeatletter
\providecommand \@ifxundefined [1]{%
 \@ifx{#1\undefined}
}%
\providecommand \@ifnum [1]{%
 \ifnum #1\expandafter \@firstoftwo
 \else \expandafter \@secondoftwo
 \fi
}%
\providecommand \@ifx [1]{%
 \ifx #1\expandafter \@firstoftwo
 \else \expandafter \@secondoftwo
 \fi
}%
\providecommand \natexlab [1]{#1}%
\providecommand \enquote  [1]{``#1''}%
\providecommand \bibnamefont  [1]{#1}%
\providecommand \bibfnamefont [1]{#1}%
\providecommand \citenamefont [1]{#1}%
\providecommand \href@noop [0]{\@secondoftwo}%
\providecommand \href [0]{\begingroup \@sanitize@url \@href}%
\providecommand \@href[1]{\@@startlink{#1}\@@href}%
\providecommand \@@href[1]{\endgroup#1\@@endlink}%
\providecommand \@sanitize@url [0]{\catcode `\\12\catcode `\$12\catcode
  `\&12\catcode `\#12\catcode `\^12\catcode `\_12\catcode `\%12\relax}%
\providecommand \@@startlink[1]{}%
\providecommand \@@endlink[0]{}%
\providecommand \url  [0]{\begingroup\@sanitize@url \@url }%
\providecommand \@url [1]{\endgroup\@href {#1}{\urlprefix }}%
\providecommand \urlprefix  [0]{URL }%
\providecommand \Eprint [0]{\href }%
\providecommand \doibase [0]{http://dx.doi.org/}%
\providecommand \selectlanguage [0]{\@gobble}%
\providecommand \bibinfo  [0]{\@secondoftwo}%
\providecommand \bibfield  [0]{\@secondoftwo}%
\providecommand \translation [1]{[#1]}%
\providecommand \BibitemOpen [0]{}%
\providecommand \bibitemStop [0]{}%
\providecommand \bibitemNoStop [0]{.\EOS\space}%
\providecommand \EOS [0]{\spacefactor3000\relax}%
\providecommand \BibitemShut  [1]{\csname bibitem#1\endcsname}%
\let\auto@bib@innerbib\@empty
\bibitem [{\citenamefont {Lai}\ and\ \citenamefont
  {T\'{e}l}(2010)}]{LaiTamasBook}%
  \BibitemOpen
  \bibfield  {author} {\bibinfo {author} {\bibfnamefont {Y.-C.}\ \bibnamefont
  {Lai}}\ and\ \bibinfo {author} {\bibfnamefont {T.}~\bibnamefont {T\'{e}l}},\
  }\href {\doibase 10.1007/978-1-4419-6987-3} {\emph {\bibinfo {title}
  {{Transient chaos: Complex dynamics in finite time scales}}}},\ \bibinfo
  {edition} {1st}\ ed.,\ edited by\ \bibinfo {editor} {\bibfnamefont {S.~S.}\
  \bibnamefont {Antman}}, \bibinfo {editor} {\bibfnamefont {J.~E.}\
  \bibnamefont {Marsden}}, \ and\ \bibinfo {editor} {\bibfnamefont
  {L.}~\bibnamefont {Sirovich}},\ Vol.\ \bibinfo {volume} {173}\ (\bibinfo
  {publisher} {Springer},\ \bibinfo {address} {New York},\ \bibinfo {year}
  {2010})\BibitemShut {NoStop}%
\bibitem [{\citenamefont {T\'{e}l}\ and\ \citenamefont {Lai}(2008)}]{Tel2008}%
  \BibitemOpen
  \bibfield  {author} {\bibinfo {author} {\bibfnamefont {T.}~\bibnamefont
  {T\'{e}l}}\ and\ \bibinfo {author} {\bibfnamefont {Y.-C.}\ \bibnamefont
  {Lai}},\ }\href {\doibase 10.1016/j.physrep.2008.01.001} {\bibfield
  {journal} {\bibinfo  {journal} {Physics Reports}\ }\textbf {\bibinfo {volume}
  {460}},\ \bibinfo {pages} {245} (\bibinfo {year} {2008})}\BibitemShut
  {NoStop}%
\bibitem [{\citenamefont {Rempel}\ and\ \citenamefont
  {Chian}(2007)}]{Rempel2007}%
  \BibitemOpen
  \bibfield  {author} {\bibinfo {author} {\bibfnamefont {E.~L.}\ \bibnamefont
  {Rempel}}\ and\ \bibinfo {author} {\bibfnamefont {A.~C.-L.}\ \bibnamefont
  {Chian}},\ }\href {\doibase 10.1103/PhysRevLett.98.014101} {\bibfield
  {journal} {\bibinfo  {journal} {Physical Review Letters}\ }\textbf {\bibinfo
  {volume} {98}},\ \bibinfo {pages} {014101} (\bibinfo {year}
  {2007})}\BibitemShut {NoStop}%
\bibitem [{\citenamefont {Taylor}\ and\ \citenamefont
  {Campbell}(2007)}]{Taylor2007}%
  \BibitemOpen
  \bibfield  {author} {\bibinfo {author} {\bibfnamefont {S.~R.}\ \bibnamefont
  {Taylor}}\ and\ \bibinfo {author} {\bibfnamefont {S.~A.}\ \bibnamefont
  {Campbell}},\ }\href {\doibase 10.1103/PhysRevE.75.046215} {\bibfield
  {journal} {\bibinfo  {journal} {Physical Review E - Statistical, Nonlinear
  and Soft Matter Physics}\ }\textbf {\bibinfo {volume} {75}},\ \bibinfo
  {pages} {046215} (\bibinfo {year} {2007})}\BibitemShut {NoStop}%
\bibitem [{Note1()}]{Note1}%
  \BibitemOpen
  \bibinfo {note} {The higher the escape time, the nearer the point is to the
  stable manifold of the invariant set, the nearer $F^{\tau /2}(\protect
  \boldsymbol {x})$ is to the chaotic saddle and nearer $F^{\tau }(\protect
  \boldsymbol {x})$ is to the unstable manifold.~\cite
  {LaiTamasBook}}\BibitemShut {NoStop}%
\bibitem [{\citenamefont {Sweet}\ \emph {et~al.}(2001)\citenamefont {Sweet},
  \citenamefont {Nusse},\ and\ \citenamefont {Yorke}}]{Sweet2001}%
  \BibitemOpen
  \bibfield  {author} {\bibinfo {author} {\bibfnamefont {D.}~\bibnamefont
  {Sweet}}, \bibinfo {author} {\bibfnamefont {H.~E.}\ \bibnamefont {Nusse}}, \
  and\ \bibinfo {author} {\bibfnamefont {J.~A.}\ \bibnamefont {Yorke}},\ }\href
  {http://prl.aps.org/abstract/PRL/v86/i11/p2261\_1} {\bibfield  {journal}
  {\bibinfo  {journal} {Physical Review Letters}\ }\textbf {\bibinfo {volume}
  {86}},\ \bibinfo {pages} {2261} (\bibinfo {year} {2001})}\BibitemShut
  {NoStop}%
\bibitem [{\citenamefont {Bollt}(2005)}]{Bollt2005}%
  \BibitemOpen
  \bibfield  {author} {\bibinfo {author} {\bibfnamefont {E.~M.}\ \bibnamefont
  {Bollt}},\ }\href@noop {} {\bibfield  {journal} {\bibinfo  {journal}
  {International Journal of Bifurcation and Chaos}\ }\textbf {\bibinfo {volume}
  {15}},\ \bibinfo {pages} {1615} (\bibinfo {year} {2005})}\BibitemShut
  {NoStop}%
\bibitem [{\citenamefont {de~Moura}\ and\ \citenamefont
  {Grebogi}(2001)}]{DeMoura2001}%
  \BibitemOpen
  \bibfield  {author} {\bibinfo {author} {\bibfnamefont {A.~P.~S.}\
  \bibnamefont {de~Moura}}\ and\ \bibinfo {author} {\bibfnamefont
  {C.}~\bibnamefont {Grebogi}},\ }\href {\doibase 10.1103/PhysRevLett.86.2778}
  {\bibfield  {journal} {\bibinfo  {journal} {Physical Review Letters}\
  }\textbf {\bibinfo {volume} {86}},\ \bibinfo {pages} {2778} (\bibinfo {year}
  {2001})}\BibitemShut {NoStop}%
\bibitem [{\citenamefont {Dellago}\ \emph {et~al.}(2002)\citenamefont
  {Dellago}, \citenamefont {Bolhuis},\ and\ \citenamefont
  {Geissler}}]{Dellago2002}%
  \BibitemOpen
  \bibfield  {author} {\bibinfo {author} {\bibfnamefont {C.}~\bibnamefont
  {Dellago}}, \bibinfo {author} {\bibfnamefont {P.}~\bibnamefont {Bolhuis}}, \
  and\ \bibinfo {author} {\bibfnamefont {P.}~\bibnamefont {Geissler}},\ }\href
  {http://dx.doi.org/10.1002/0471231509.ch1}
  {\bibfield  {journal} {\bibinfo  {journal} {Advances in Chemical Physics}\ }\textbf{\bibinfo {volume} {123}}
  (\bibinfo {year} {2002})}\BibitemShut {NoStop}%
\bibitem [{\citenamefont {Leit\~{a}o}\ \emph {et~al.}(2014)\citenamefont
  {Leit\~{a}o}, \citenamefont {Lopes},\ and\ \citenamefont
  {Altmann}}]{Leitao2014}%
  \BibitemOpen
  \bibfield  {author} {\bibinfo {author} {\bibfnamefont {J.~C.}\ \bibnamefont
  {Leit\~{a}o}}, \bibinfo {author} {\bibfnamefont {J.~M. V.~P.}\ \bibnamefont
  {Lopes}}, \ and\ \bibinfo {author} {\bibfnamefont {E.~G.}\ \bibnamefont
  {Altmann}},\ }\href {\doibase 10.1103/PhysRevE.90.052916} {\bibfield
  {journal} {\bibinfo  {journal} {Physical Review E}\ }\textbf {\bibinfo
  {volume} {90}},\ \bibinfo {pages} {052916} (\bibinfo {year}
  {2014})}\BibitemShut {NoStop}%
\bibitem [{\citenamefont {Leit\~{a}o}\ \emph {et~al.}(2013)\citenamefont
  {Leit\~{a}o}, \citenamefont {Lopes},\ and\ \citenamefont
  {Altmann}}]{Leitao2013}%
  \BibitemOpen
  \bibfield  {author} {\bibinfo {author} {\bibfnamefont {J.~C.}\ \bibnamefont
  {Leit\~{a}o}}, \bibinfo {author} {\bibfnamefont {J.~M. V.~P.}\ \bibnamefont
  {Lopes}}, \ and\ \bibinfo {author} {\bibfnamefont {E.~G.}\ \bibnamefont
  {Altmann}},\ }\href {\doibase 10.1103/PhysRevLett.110.220601} {\bibfield
  {journal} {\bibinfo  {journal} {Physical Review Letters}\ }\textbf {\bibinfo
  {volume} {110}},\ \bibinfo {pages} {220601} (\bibinfo {year}
  {2013})}\BibitemShut {NoStop}%
\bibitem [{Note2()}]{Note2}%
  \BibitemOpen
  \bibinfo {note} {In Ref.~\cite {Sweet2001} the distribution is constructed by
  considering $\delta =10^{-s}$ with $s$ chosen randomly from a uniform
  distribution in $[\protect \qopname \relax o{log}_{10}(\delta _{min}),
  \protect \qopname \relax o{log}_{10}(\delta _{max})]$}\BibitemShut {NoStop}%
\bibitem [{\citenamefont {Gr\"{u}nwald}\ \emph {et~al.}(2008)\citenamefont
  {Gr\"{u}nwald}, \citenamefont {Dellago},\ and\ \citenamefont
  {Geissler}}]{Grunwald2008}%
  \BibitemOpen
  \bibfield  {author} {\bibinfo {author} {\bibfnamefont {M.}~\bibnamefont
  {Gr\"{u}nwald}}, \bibinfo {author} {\bibfnamefont {C.}~\bibnamefont
  {Dellago}}, \ and\ \bibinfo {author} {\bibfnamefont {P.~L.}\ \bibnamefont
  {Geissler}},\ }\href {http://link.aip.org/link/?JCPSA6/129/194101/1}
  {\bibfield  {journal} {\bibinfo  {journal} {The Journal of chemical physics}\
  }\textbf {\bibinfo {volume} {129}},\ \bibinfo {pages} {194101} (\bibinfo
  {year} {2008})}\BibitemShut {NoStop}%
\bibitem [{Note3()}]{Note3}%
  \BibitemOpen
  \bibinfo {note} {To compute $\tau $, the map has to be iterated $\tau $ times
  and thus the number of map iterations needed to reach $\tau _{\protect
  \qopname \relax m{max}}$ is of the order of $\DOTSB \sum@ \slimits@ _{\tau
  =1}^{\tau _{\protect \qopname \relax m{max}}} C(\tau )\tau $.}\BibitemShut
  {Stop}%
\bibitem [{\citenamefont {Kuptsov}\ and\ \citenamefont
  {Parlitz}(1990)}]{Kuptsov2012}%
  \BibitemOpen
  \bibfield  {author} {\bibinfo {author} {\bibfnamefont {P.V.}\ \bibnamefont
  {Kuptsov}}\ and\ \bibinfo {author} {\bibfnamefont {U.}~\bibnamefont
  {Parlitz}},\ }\href {\doibase 10.1007/s00332-012-9126-5}
 {\bibfield
  {journal} {\bibinfo  {journal} {Journal of Nonlinear Science}\ }\textbf {\bibinfo {volume}
  {22}},\ \bibinfo {pages} {727} (\bibinfo {year} {2012})}\BibitemShut
  {NoStop}%
\end{thebibliography}

\end{document}